%% file: main.tex
\documentclass[sigconf]{acmart}

\input{macros}

\usepackage[edges]{forest}
\usepackage{microtype}

\definechangesauthor[color=blue]{li}

\renewcommand\footnotetextcopyrightpermission[1]{} 

\begin{document}


\title{Demystifying the Characteristics for Smart Contract Upgrades}


\author{Ye Liu}
\email{yeliu@smu.edu.sg}
\affiliation{%
	\institution{Singapore Management University}
	\country{Singapore}
}
\author{Shuo Li}
\email{lishuo19@otcaix.iscas.ac.cn}
\affiliation{%
	\institution{Institute of Software, Chinese Academy of Sciences}
	\country{China}
}

\author{Xiuheng Wu}
\email{xiuheng.wu@ntu.edu.sg}
\affiliation{%
	\institution{Nanyang Technological University}
	\country{Singapore}
}

\author{Yi Li}
\email{yi\_li@ntu.edu.sg}
\orcid{0000-0003-4562-8208}
\affiliation{%
	\institution{Nanyang Technological University}
	\country{Singapore}
}

\author{Zhiyang Chen}
\email{zhiychen@cs.toronto.edu}
\affiliation{%
	\institution{University of Toronto}
	\country{Canada}
}

\author{David Lo}
\email{davidlo@smu.edu.sg}
\affiliation{%
	\institution{Singapore Management University}
	\country{Singapore}
}


\begin{abstract}
Upgradable smart contracts play an important role in the decentralized application ecosystem, to
support routine maintenance, security patching, and feature additions.
In this paper, we conduct an empirical study on proxy-based upgradable smart contracts to
understand the characteristics of contract upgrading.
Through our study on 57,118 open source proxy contracts, we found that 583 contracts have ever been
upgraded on Ethereum, involving 973 unique implementation contract versions.
The results show that developers often intend to improve usability of contracts if upgrading, where
functionality addition and update are the most frequent \code{upgrade intentions}.
We investigated the practical impacts of contract upgrades, e.g., \code{breaking changes} causing
compatibility issues, \code{storage collisions} and \code{initialization risks} leading to security
vulnerabilities.
The results demonstrate that there are 4,334 ABI breaking changes due to the upgrades of 276
proxies, causing real-world broken usages within 584 transactions witnessed by the blockchain;
36 contract upgrades had storage collisions and five proxies with 59 implementation contracts are
vulnerable to initialization attacks.
\end{abstract}

\maketitle

\input{intro}

\input{background}
\input{design}
\input{detection}

\input{study}
\input{discuss}
\input{related}
\input{conclusion}

\bibliographystyle{ACM-Reference-Format}
\bibliography{references}

\end{document}

%% file: macros.tex
\usepackage{graphicx}

\usepackage{textcomp}
\usepackage{xcolor}
\def\BibTeX{{\rm B\kern-.05em{\sc i\kern-.025em b}\kern-.08em
    T\kern-.1667em\lower.7ex\hbox{E}\kern-.125emX}}
\usepackage[utf8]{inputenc}

\usepackage{xspace}
\usepackage{xcolor}
\usepackage{graphicx}
\usepackage[normalem]{ulem} 
\usepackage{enumitem}
\usepackage{microtype}
\usepackage{hyphenat}
\usepackage{booktabs}
\usepackage{multirow}
\usepackage{makecell}
\usepackage{ragged2e}
\usepackage{longtable}
\usepackage{diagbox}
\usepackage{caption}
\usepackage{subcaption}
\usepackage{wrapfig}
\usepackage{listings}

\usepackage[many]{tcolorbox}
\usepackage{lipsum}

\usepackage{bussproofs}
\usepackage{stackengine}

\usepackage{pifont}
\usepackage{tikz}
\usetikzlibrary{calc}
\usetikzlibrary{shapes.geometric}
\usetikzlibrary{decorations.pathreplacing}
\usetikzlibrary{positioning}
\usetikzlibrary{calc}
\usetikzlibrary{arrows}
\usetikzlibrary{shadows}

\usepackage{balance}
\usepackage{datetime} 

\usepackage[capitalize]{cleveref}
\usepackage{hyperref}

\usepackage{algorithm}
\usepackage[noend]{algpseudocode}

\usepackage{venndiagram}

\crefname{section}{Sect.}{Sects.}
\Crefname{section}{Section}{Sections}

\newcommand{\XSpace}[1]{}
\newcommand{\XComment}[1]{}
\usepackage{todonotes} 

\newcommand{\DefMacro}[2]{\expandafter\newcommand\csname rmk-#1\endcsname{#2}}
\newcommand{\UseMacro}[1]{\csname rmk-#1\endcsname}

\renewcommand{\paragraph}[1]{\vskip 0.05in \noindent\textbf{#1.}}

\newcommand{\InputWithSpace}[1]{\bgroup\def\arraystretch{1.1}\input{#1}\egroup}
\newcommand{\Code}[1]{{\ifmmode{\mathtt{#1}}\else$\mathtt{#1}$\fi}}
\newcommand{\CodeIn}[1]{{\ifmmode{\mathtt{#1}}\else$\mathtt{#1}$\fi}}

\newcommand{\tool}{\textsc{UpCon}\xspace}

\definecolor{silver}{RGB}{192,192,192}

\newcolumntype{R}[1]{>{\RaggedLeft\arraybackslash}p{#1}}
\newcolumntype{L}[1]{>{\RaggedRight\arraybackslash}p{#1}}

\DefMacro{NumOfSubjectsReduceOver90Percent}{seven}

\newtcolorbox{rqbox}[1]{%
    tikznode boxed title,
    enhanced,
    arc=0mm,
    boxrule=0.5pt,
    interior style={white},
    attach boxed title to top center= {yshift=-\tcboxedtitleheight/2},
    colbacktitle=white,coltitle=black,
    boxed title style={size=normal,colframe=white,boxrule=0pt},
    title=\textbf{Answer to }{\textbf{#1}}}

\definecolor{codegreen}{rgb}{0,0.6,0}
\definecolor{codegray}{rgb}{0.5,0.5,0.5}
\definecolor{codepurple}{rgb}{0.58,0,0.82}
\definecolor{backcolour}{rgb}{0.95,0.95,0.92}

\usepackage{alphabeta}

\usepackage[group-separator={,}]{siunitx}

\usepackage{graphicx}
\usepackage{booktabs}
\usepackage{multirow}
\usepackage[flushleft]{threeparttable}
\usepackage{xcolor}
\usepackage{syntax}

\usepackage[frozencache,cachedir=.]{minted}

\usepackage{changes}
\usepackage{blkarray, bigstrut}
\newcommand{\setBAcolsep}[1]{\BA@colsep=#1}
\usepackage{algorithm}
\usepackage[noend]{algpseudocode}
\usepackage{algorithmicx}

\usepackage{pgfplotstable}

\usepackage{amsmath,amssymb,amsfonts}

\usepackage{soul}

\usepackage{bussproofs}

\usepackage{colortbl}

\input{solidity-highlighting}

\newcommand{\website}{\url{https://sites.google.com/view/upcon-home/}}

\usepackage{tikz}
\usetikzlibrary{automata, positioning, arrows, shapes}

\tikzstyle{block} = [rectangle, draw, text width=5em,
text centered, rounded corners,
minimum height=4em, node distance=3cm]
\tikzstyle{line} = [draw, -latex']
\tikzstyle{cloud} = [draw, ellipse, node distance=2.5cm, minimum height=2em]
\tikzstyle{blank} = [node distance=1cm]

\renewcommand{\#}{\texttt{\symbol{`\#}}}
\usepackage{bussproofs}

\usepackage{listings,multicol}  
\usepackage{filecontents}       

\newsavebox{\mybox}

\newcommand{\code}[1]{\emph{#1}}

%% file: solidity-highlighting.tex
\usepackage{listings, xcolor}

\definecolor{verylightgray}{rgb}{.97,.97,.97}

\lstdefinelanguage{Solidity}{
	keywords=[1]{anonymous, assembly, assert, balance, break, call, callcode, case, catch, class, constant, continue, constructor, contract, debugger, default, delegatecall, delete, do, else, emit, event, experimental, export, external, false, finally, for, function, gas, if, implements, import, in, indexed, instanceof, interface, internal, is, length, library, log0, log1, log2, log3, log4, memory, modifier, new, payable, pragma, private, protected, public, pure, push, require, return, returns, revert, selfdestruct, send, solidity, storage, struct, suicide, super, switch, then, this, throw, true, try, typeof, using, value, view, while, with, addmod, ecrecover, keccak256, mulmod, ripemd160, sha256, sha3}, 
	keywordstyle=[1]\color{blue}\bfseries,
	keywords=[2]{address, bool, byte, bytes, bytes1, bytes2, bytes3, bytes4, bytes5, bytes6, bytes7, bytes8, bytes9, bytes10, bytes11, bytes12, bytes13, bytes14, bytes15, bytes16, bytes17, bytes18, bytes19, bytes20, bytes21, bytes22, bytes23, bytes24, bytes25, bytes26, bytes27, bytes28, bytes29, bytes30, bytes31, bytes32, enum, int, int8, int16, int24, int32, int40, int48, int56, int64, int72, int80, int88, int96, int104, int112, int120, int128, int136, int144, int152, int160, int168, int176, int184, int192, int200, int208, int216, int224, int232, int240, int248, int256, mapping, string, uint, uint8, uint16, uint24, uint32, uint40, uint48, uint56, uint64, uint72, uint80, uint88, uint96, uint104, uint112, uint120, uint128, uint136, uint144, uint152, uint160, uint168, uint176, uint184, uint192, uint200, uint208, uint216, uint224, uint232, uint240, uint248, uint256, var, void, ether, finney, szabo, wei, days, hours, minutes, seconds, weeks, years},	
	keywordstyle=[2]\color{teal}\bfseries,
	keywords=[3]{block, blockhash, coinbase, difficulty, gaslimit, number, timestamp, msg, data, gas, sender, sig, value, now, tx, gasprice, origin},	
	keywordstyle=[3]\color{violet}\bfseries,
	identifierstyle=\color{black},
	sensitive=false,
	comment=[l]{//},
	morecomment=[s]{/*}{*/},
	commentstyle=\color{gray}\ttfamily,
	stringstyle=\color{red}\ttfamily,
	morestring=[b]',
	morestring=[b]"
}

%% file: intro.tex
\section{Introduction}
Smart contracts are programs deployed and executed on blockchain platforms, automating trustless
transactions between users.
The \emph{upgradable smart contract} is a popular design paradigm to develop smart contracts that
can be upgraded transparently.
Although contract codes remain immutable once deployed to blockchains, upgradability design
patterns, e.g., proxy patterns~\cite{eip1538,eip1822,eip1967,eip2535,eip3561,eip897} and data
separation patterns~\cite{antipatterns}, allow flexible modifications without altering the original
contract's address.
Upgradable smart contracts have been widely used in popular decentralized applications, e.g.,
Uniswap~\cite{uniswap} for building financial applications and OpenSea~\cite{opensea} for supporting
NFT marketplaces.
To support development of upgradable smart contracts,
the leading third-party contract library vendor, OpenZeppelin~\cite{OpenZeppelin}, has created many
reference implementations for different upgradability standards~\cite{eip1538, eip1822,eip2535}.

However, implementing upgradable smart contracts introduces challenges.
While upgradable contracts are one of the most effective ways to resolve bugs after deployment,
upgrading them can increase the complexity of the existing smart contract architecture and may
inadvertently introduce new bugs.
For example, the recent security incidents involving Pike Finance illustrate the risks associated
with using upgradable smart contracts~\cite{Pikehack}.
The Pike Finance team used upgradable smart contracts in its cross-chain transfer protocol.
On April 26, 2024, Pike Finance first suffered a hack due to a logic bug, and the attackers stole
about \$300,000 from the protocol.
In response to this initial hack, the team upgraded the contract to fix the bug with the inclusion of inheriting an additional dependency code.
This dependency introduced new variables
that altered the contract's storage layout, causing a \emph{storage collision} problem with the previous contract version.
One effect was that a contract state variable ``initialized'' was no longer accessible, behaving as
if it had not been initialized.
Subsequently, another attacker exploited this vulnerability to upgrade the contract for the second time with a malicious version.
By doing so, he gained administrator privileges and was able to access the funds deposited into Pike Finance.
This allowed the attacker to carry out the second, \$1.6 million theft from the Pike protocol.
Apart from bug fixes, adding new functionalities or even improving contract features in upgrades
could also introduce severe vulnerabilities~\cite{euler, uranium,nowswap, nomad}.
For example, Euler Finance was hacked due to the addition of a vulnerable function
\code{donateToReserves}~\cite{euler}, resulting in a loss of \$200 million in 2023.
Despite the huge impact, it is still very challenging to develop effective countermeasures to avoid
the above attacks, without a good understanding of real-world smart contract upgrades.

Smart contract upgrades have not attracted much research attention.
There is a large body of analysis techniques for standalone smart contracts.
To detect security vulnerabilities, methods like static analysis~\cite{feist2019slither,tsankov2018securify}, fuzzing~\cite{2018contractfuzzer,nguyen2020sfuzz}, and symbolic execution~\cite{mossberg2019manticore, Mythril} have been proposed and have achieved great success.
Additionally, formal verification approach~\cite{wang2019formal, permenev2020verx, liu2020towards} have also been applied to establish the correctness of smart contracts in a more systematic way.
On the one hand, not only for bug fixes, upgradable smart contracts get more important to mitigate real-world attacks. \citeauthor{chen2024demystifying}~\cite{chen2024demystifying} pointed out that developers can enforce invariant guards as patches for effective defense against attacks via upgradable smart contracts.
They studied 42 victim contracts in 27 real-world exploits that cumulatively resulted in financial losses exceeding \$2 billion.
Particularly, we found 15 of these victim contracts are proxy-based upgradable smart contracts and the patches can be easily implemented by simple contract upgrades.
On the other hand,
there are only a few empirical studies on the summarisation of practical upgrade mechanisms, i.e., upgradability patterns for smart contracts~\cite{antonino2022specification,salehi2022not,xiaofan2024, bodell2023proxy}.
For upgrade issues, \citeauthor{bodell2023proxy}~\cite{bodell2023proxy} mainly studied the insufficient compatibility check issues within the \code{setter} functions of proxy contracts.
\citeauthor{sorensen2024towards}~\cite{sorensen2024towards} proposed a Coq~\cite{huet1997coq} based formal verification tool to protect smart contract upgrades to avoid faulty updates.
Yet, it remains unknown of the characteristics for real-world contract upgrades, namely what and why contracts upgrade, and how they impact client-specific compatibility and its own security.

In this empirical study,
we aim to fill this research gap and characterize the upgrading behaviors for smart contracts.
First, we investigated the contents of upgrades, specifically analyzing the code changes between
the original and the updated versions, to understand the modifications made.
Then, we systematically studied the underlying intentions for these upgrades to capture the
motivations behind contract upgrades.
Finally, we further investigated the impacts of contract upgrades in terms of their compatibility
and security implications.
We discovered many \emph{breaking changes} and two common upgrade-related vulnerabilities including
\emph{storage collision} and \emph{initialization risks} that apparently affect the data integrity
of smart contracts.

We implemented the aforementioned analysis modules in a framework called \tool.
Particularly, we designed a comprehensive upgrading intention taxonomy for real-world smart
contract upgrades, to ensure the reliability of manual analysis results.
We systematically collected 57,118 proxy contracts from 8 blockchains and found that 583 proxies have ever
been upgraded on Ethereum, with 973 unique implementation contract verions.
The most common code changes and upgrading intentions are about {feature additions and updates} to improve contract usability.
We found in total 4,334 ABI {(application binary interface)} breaking changes of 276 proxies.
Furthermore, we examined the past blockchain transactions and found broken usages within 584 transactions caused by these ABI breaking changes.
For security issues, we identified 36 storage collision cases while 5 proxies and 59 implementation
contracts have initialization risks.

The main contributions of this paper are summarized below:
 \begin{itemize}[leftmargin=*]
 	\item We propose and implement~\tool, a novel analysis framework for characterizing smart contract upgrades.
 To facilitate this analysis, we customized a comprehensive upgrade intention taxonomy for smart contracts.
 	\item We evaluate \tool on a large-scale study of 44,282 proxies on Ethereum. Our research uncovers critical insights into upgradable smart contracts, highlighting the essential need for developers to address compatibility and security concerns during upgrades and calling for more focused academic research in this area.
 	\item We provide open access to the prototype tools, dataset, and analysis results utilized in this study at \website, facilitating further research and development in the community.
 \end{itemize}


\paragraph{Organization}
The rest of the paper is organized as follows.
\Cref{sec:background} presents the background about upgradable smart contracts.
\Cref{sec:methodology} illustrates our analysis framework~\tool, and the data collection process.
\Cref{sec:characteristics} presents the findings of upgrade changes, intentions, compatibility and security issues.
\Cref{sec:discussion} discusses our research implications and the validity threats.
The related works are compared in~\cref{sec:related} and we conclude this paper in~\cref{sec:conclusion}.

%% file: background.tex
\section{Background}
\label{sec:background}
The upgradable smart contract is a crucial advancement in smart contract development, addressing
the limitations brought by contract immutability.
Traditionally, once a smart contract is deployed onto a blockchain, its code cannot be altered
anymore.
This immutability ensures security and trustlessness but poses challenges for developers who need
to fix bugs or add new features to the contract.

\paragraph{Benefits}
Upgradable smart contracts solve this problem by allowing modifications without altering the
original contract's address or disrupting the network's trust.
Various design patterns enable upgradability,
such as \emph{proxy pattern}, data separation pattern, and metamorphic
contracts~\cite{xiaofan2024}. The most commonly used one is the {proxy pattern}, where a proxy
contract delegates calls to an implementation contract. This setup allows the implementations of
upgradable smart contracts to be effectively replaced in certain \emph{setter} functions,
transparently upgrading contract's logic that does not change the original proxy contract address.
With contract upgrades,  developers can perform bug fixes, introduce new functionality, or improve existing contract features.

\begin{figure}[t]
	\centering
	\includegraphics[width=.9\linewidth]{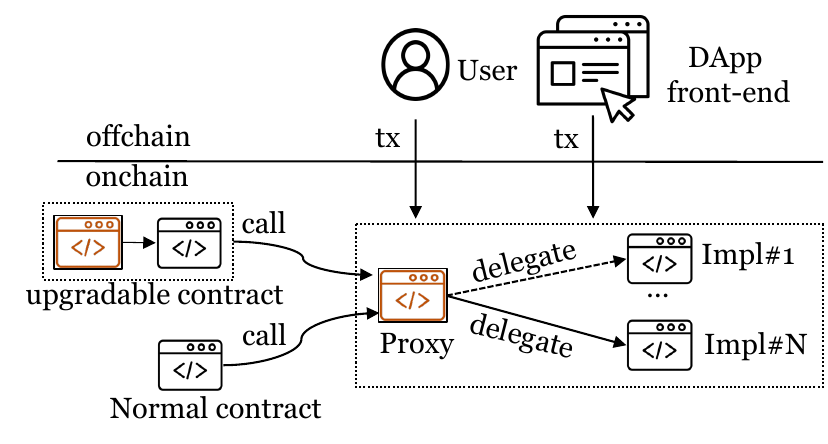}
	\caption{The usage scenarios of upgradable smart contracts.}
	\label{fig:use}
\end{figure}

\paragraph{Use Scenarios}
\Cref{fig:use} demonstrates the usage scenarios of proxy-based upgradable smart contracts.
An upgradable smart contract consists of a proxy contract and one or more implementation contracts.
Upgradable contracts can be directly accessed from other smart contracts onchain, while
externally-owned user accounts and offchain DApps front-ends can send transactions to the
blockchain to trigger contract execution.
All uses of upgradable smart contracts should meet its Application Binary Interface (ABI)
specifications.
Although invalid usages will be discarded by the blockchain, it may affect the normal execution of
the related business logic, bringing up non-refundable transaction costs.
When proxy contracts replace old implementations with new ones, breaking changes might be
introduced, increasing the possibility of invalid contract usages (we study this phenomenon in
detail in~\cref{sec:rq3}).

\paragraph{Risks}
There are two main risks centered around upgradable smart contracts.
First, implementing upgradable contracts increases the complexity of the system, which may introduce new bugs~\cite{euler, uranium, nomad,nowswap}.
Second, it is challenging for users to trust that upgrades will be beneficial and not malicious {because it is often the contract deployers that hold the centralized power to control the upgrade process}.
Unlike standalone contracts, thorough testing or audit just before contract deployment are not
sufficient since the design requirement and runtime environment may evolve and should be
verified~\cite{sorensen2024towards} after each contract upgrade.


To mitigate such risks, in this empirical study, we characterize real-world smart contract
upgrades, which differs from the existing studies~\cite{antonino2022specification, bodell2023proxy,
salehi2022not,xiaofan2024, bui2021evaluating} focusing only on contract upgradability.
We hope this study could shed light into the complexity and risk of contract upgrades and advance towards the responsible adoption of upgradable smart contracts.

%% file: design.tex
\section{Methodology}\label{sec:methodology}
\begin{figure}[t]
	\centering
   \includegraphics[width=.99\columnwidth]{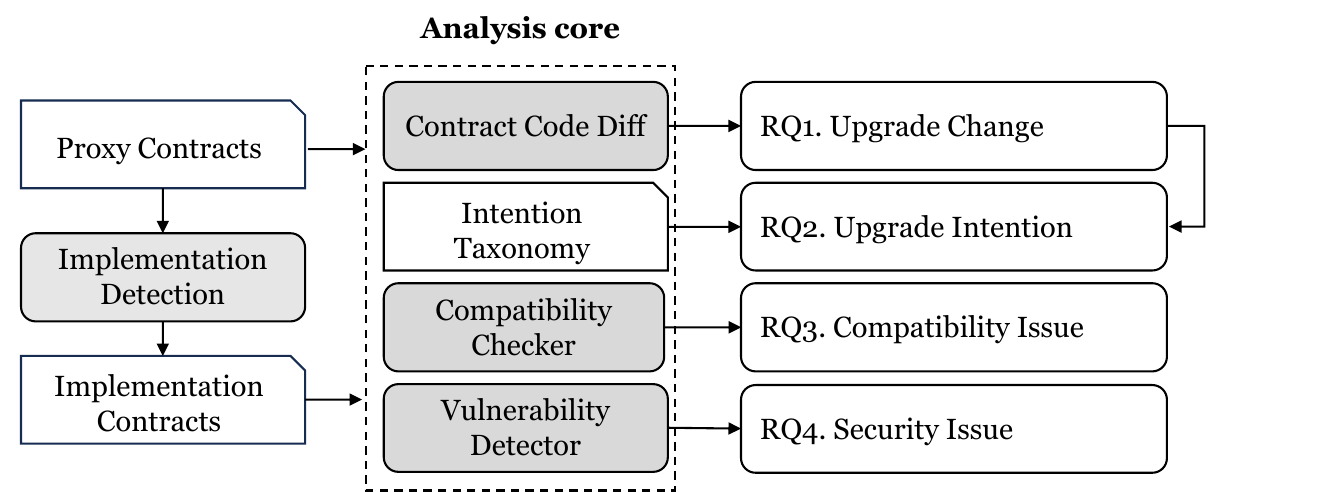}
   \caption{The \tool framework for contract upgrades.}
   \label{fig:overview}
\end{figure}

\Cref{fig:overview} shows our framework, \tool, for smart contract upgrade analysis.
We study the widely-used proxy-based upgradable smart contracts, and propose a highly effective
detection approach to collect all its historical implementation contract versions (find more
details in~\cref{sec:detection}).
Incorporated with an analysis toolbox comprising contract code differencing, a comprehensive
upgrade intention taxonomy, effective compatibility checker and vulnerability detector,
\tool aims to answer the following four research questions:
\begin{itemize}[leftmargin=*]
	\item RQ1. (\textbf{Upgrade Change}) What code changes are introduced in contract upgrades?
	\item RQ2. (\textbf{Upgrade Intention})  Why are smart contracts upgraded?
	\item RQ3. (\textbf{Compatibility Issue}) What client-specific compatibility issues are caused by contract upgrading?
	\item RQ4. (\textbf{Security Issue}) Can~\tool find security issues in upgradable smart contracts?
\end{itemize}

We answer RQ1 and RQ2 in~\cref{sec:rq1} and~\cref{sec:rq2}, respectively, to address what and why changes are introduced in smart contract upgrades.
We answer RQ3 in~\cref{sec:rq3} to reveal whether smart contract upgrades introduce ABI breaking changes, leading to client-specific compatibility issues.
RQ4 was answered in~\cref{sec:rq4} and we focus on the detection of \emph{storage collision} and \emph{initialization risks} that are critical to the integrity of upgradable smart contracts.

\subsection{Overview of Proxy Contracts}
\begin{table}[t]\centering
	\caption{The statistics of contracts from different blockchains.}\label{tab: }
	\centering
	\begin{tabular}{lrrc}\toprule
		Chain &\#Contract &\#Proxy &\#Proxy/\#Contract \\\midrule
		Ethereum &692,369 &44,282 &0.064 \\
		OP Mainnet &21,215 &2,542 &0.120 \\
		BNB Smart Chain &486,922 &1,904 &0.004 \\
		Polygon &178,630 &3,001 &0.017 \\
		Fantom Opera &41,745 &310 &0.007 \\
		Base &19,673 &1,880 &0.096 \\
		Arbitrum One &56,061 &1,161 &0.021 \\
		Avalanche C-Chain &49,029 &2,038 &0.042 \\
		\midrule
		Overall &1,545,644 &57,118 &0.037 \\
		\bottomrule
	\end{tabular}
	\label{fig:proxy}
\end{table}

We systematically crawled all open source smart contracts from Etherscan~\cite{Etherscan}.
These contracts come from different blockchains and some of them have also been labeled as proxy
contracts according to the heuristic-based recognition
rule,\footnote{\url{https://info.etherscan.com/what-is-proxy-contract/}} where Etherscan
specifically looks for identifiable bytecode patterns, e.g., \code{delegatecall} opcode, associated
with well-known proxy contract implementations.

\Cref{fig:proxy} shows the statistics of open source contracts from multiple blockchains.
Ethereum has the most contracts (692,369) and 6.4\% of them are proxy contracts (44,282), accounting for 77.5\%  of such contracts (57,118) in total.
Although BNB Smart Chain and Polygon have the second and third most contracts but they contain quite smaller number of proxy contracts compared with Ethereum.
In this work, we give priority to those proxy contracts having sufficient historical implementation
contracts for creating a large benchmark dataset full of contract upgrades.
Hence, we only investigated the 44,282 proxy contracts from Ethereum, because we believe reliable
upgrade findings could hardly be established from the smaller set of proxy contracts from other
blockchains.
In the next section, we detail the approach to identify all the corresponding logic implementations
for the proxy contracts studied.

%

%% file: detection.tex
\subsection{ Implementation Detection}
\label{sec:detection}
\paragraph{Proxy Patterns}
\citeauthor{bodell2023proxy}~\cite{bodell2023proxy} systematically investigated proxy-based upgradable smart contracts and finally identified 11 proxy patterns under seven categories.
Only two proxy patterns (EIP-1538 and EIP-2535) belong to the category of the so-called multiple
implementations, where a proxy could own two or more implementation contracts at the same time and
each implementation usually deals with a minimal set of contract functionality.
Proxies of the remaining nine patterns can only have a single working implementation contract at
any time.
For most of such proxies, the addresses of implementation contracts are queryable through ABI
functions, e.g.,  \code{implementation() public view returns(address)} in many proxy contracts
written in Solidity~\cite{solidity}.
To ensure reliable detection of implementation contracts, we only consider those proxy contracts
exposing the ``implementation()'' function, where the results in~\cref{sec:proxyimpl} indicate this
setting could cover most proxies.

\paragraph{Detection algorithm}
\Cref{algo:impldetect} illustrates our algorithm to effectively detect the addresses of implementation contracts and the corresponding block number for the contract upgrades.
Given a proxy contract address, \emph{proxy}, and the latest block number of blockchains,
\emph{latestBlockNo}, we invoke a binary search procedure
(Lines~\ref{line:binarysearch:begin}-\ref{line:binarysearch:end}) to systematically retrieve all
historical implementation contracts and the exact blocks where the contract upgrade happened.
Specifically, we search the whole blockchain history for new implementation contracts by narrowing
down search space with a divide-and-conquer strategy
(Lines~\ref{line:narrow:begin}-\ref{line:narrow:end}).
When blocks are exhausted, we find the exact proxy upgrading block for an implementation contract and record it into our results (Line~\ref{line:add}).
To this end, we can successfully obtain all implementation contracts and trace the historical
upgrading activities.

Note that the correctness of~\cref{algo:impldetect} is established under the assumption that proxies will not use a previously abandoned implementation contracts and there is at most one upgrading records in a block.
By using~\cref{algo:impldetect}, we can achieve $O(log\; N)$ in terms of time complexity, which is much faster than the existing implementation search approach employed in~\cite{bodell2023proxy} that would iteratively query all the historic transactions having $O(N)$ complexity.
In our experiment,
\cref{algo:impldetect} usually takes no more than 50 queries to recognize all implementation contracts and corresponding upgrading records for proxies on Ethereum that has around 19.4 million blocks as of March 2024.
The other advantage of~\cref{algo:impldetect} is that we can recover exact contract upgrading activities, independent of concrete upgrade mechanisms, i.e., diverse \emph{setter} functions, enforced in the proxies.

\begin{algorithm}[t]
	\caption{\textsc{ImplDetect}$(proxy, latestBlockNo)$}
	\begin{algorithmic}[1]\small
		\State $Results \gets \emptyset$ \Comment{a set of implementation contracts and exact upgrading blocks, which is updated inside \textsc{BinarySearch}.}
		\Procedure{BinarySearch}{$from$, $leftImpl$,  $end$, $rightImpl$} \label{line:binarysearch:begin}

			\If{from == end -1}
					\State $Results = Results \cup \{(rightImpl,end)\}$ \label{line:add} \label{line:update}
			\Else
					\State Let $mid = (from + end)/2$ \label{line:narrow:begin}
					\State $midImpl \gets$ \textsc{Query}$(proxy, ``implementation()", mid)$
					\If{$midImpl \neq leftIml$}
					\State \textsc{BinarySearch}(from,  leftImpl, mid,midImpl)
					\EndIf
					\If{$midImpl \neq rightImpl$}
					\State \textsc{BinarySearch}( mid,  midImpl, end, rightImpl)
					\EndIf  \label{line:narrow:end}
			\EndIf
		\EndProcedure \label{line:binarysearch:end}

		\State $nullImpl \gets  Null$
		\State $curImpl \gets$ \textsc{Query}$(proxy, "implementation()", latestBlockNo)$
		\State \textsc{BinarySearch}$(0,nullImpl, latestBlockNo,  curImpl)$
		\State \Return $Results$
	\end{algorithmic}
	\label{algo:impldetect}
\end{algorithm}

\subsection{Statistics of Contract Upgrades}
\label{sec:proxyimpl}
\begin{figure}[t]
	\centering
	\includegraphics[width=\columnwidth]{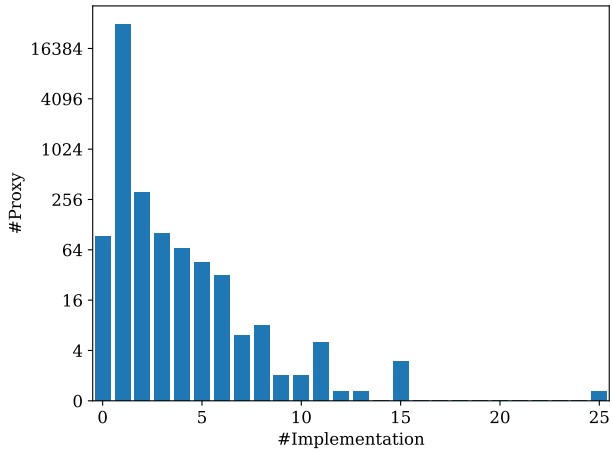}
	\caption{The distribution of proxy contracts having different implementations on Ethereum. }
	\label{fig:frequency}
\end{figure}

\begin{figure}[t]
	\includegraphics[width=\columnwidth]{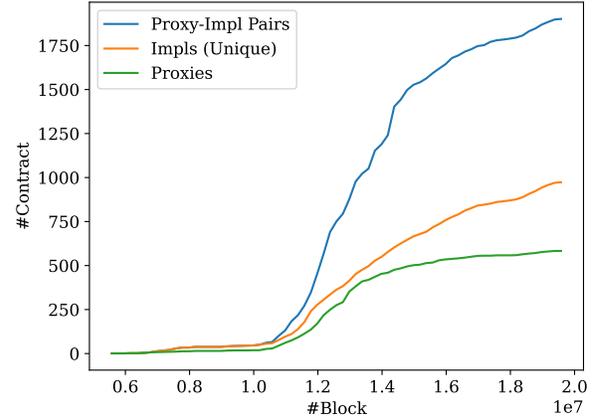}
	\caption{The number of upgraded proxies and their implementation contracts at different block checkpoints.}
	\label{fig:distribution}
\end{figure}

We used the Infura API\footnote{https://www.infura.io/} to implement \textsc{Query} to fetch the
implementation contract addresses that are persistently stored within all historical contract state
snapshots in the archive nodes.
We ran~\cref{algo:impldetect} on all 44,282 proxy contracts from Ethereum.
As a result, we successfully obtained implementation contract addresses for 32,186 proxy contracts
(proxies), accounting for 72.7\,\% in total.

\Cref{fig:frequency} shows the distribution of proxies having different number of implementations.
The x-axis categorizes these proxies according to their number of implementation contracts used.
Among 32,186 proxy contracts studied, the maximum number of implementation contracts used by proxies is 25.
Ninety four proxies (0.29\,\%) have zero implementation, which means they are uninitialized, and
31,509 proxies (97.90\,\%) have been using only one implementation.
The remaining 583 proxies (1.81\,\%) have ever upgraded its implementations, involving 1,901 proxy-implementation pairs and 973 unique implementation contracts in total.

\Cref{fig:distribution} demonstrates how these 583 proxies evolved on Ethereum with time passing.
\Cref{fig:distribution} presents the number of proxy-implementation pairs \emph{\#Pair}, the count of unique implementation contracts (\emph{\#Impl}) and proxy contracts (\#Proxy).
Overall, each proxy has three more implementation contracts used on average.
The first proxy contract was deployed on May 8, 2018 while the last proxy contract was deployed on Mar 5, 2024.
From the significant difference between \emph{\#Pair} and \emph{\#Impl},
there are apparently a considerable number of implementation contracts shared by multiple proxies.
Moreover, \emph{\#Impl} had been quite close to \emph{\#Proxy} until 1.4 million blocks, i.e., January, 2022.
Afterwards, implementation contracts are created much more than that of proxy contracts.
One of the possible reasons may be that 2022 was the biggest year ever for crypto theft with \$3.7 billion stolen~\cite{chainalysis}, raising up the awareness of active contract upgrades for bug fixes.
%

%% file: study.tex
\section{Characteristics of Smart Contract Upgrades}
\label{sec:characteristics}

\input{RQ1}
\input{RQ2}

\label{sec:issue}
\input{RQ3}

\input{RQ4}

%% file: RQ1.tex
\begin{table}[t]\centering
	\caption{Code changes of contract upgrades.}\label{tab:codechange}
	\small
	\begin{tabular}{lrrrr}\toprule
		Syntax structure &Insert &Delete &Update &Move \\\midrule
		contract &533 &471 &172 &671 \\
		state variable &535 &379 &309 &150 \\
		event &159 &492 &29 &48 \\
		modifier &20 &27 &278 &5 \\
		function &3932 &2984 &2481 &965 \\
		parameter &631 &385 &564 &636 \\
		modifier invocation &48 &23 &1763 &12 \\
		assembly statement &1 &227 &1 &0 \\
		expression statement &2714 &2880 &292 &1101 \\
		if statement &691 &1217 &58 &75 \\
		for statement &13 &7 &3 &2 \\
		while statement &2 &0 &0 &0 \\
		revert statement &5 &1 &2 &0 \\
		try statement &1 &0 &0 &0 \\
		emit statement &271 &130 &38 &136 \\
		return statement &142 &40 &177 &10 \\
		\midrule
		Overall &9698 &9263 &6167 &3811 \\
		\bottomrule
	\end{tabular}
\end{table}
\subsection{RQ1. Upgrade Change}
\label{sec:rq1}
\paragraph{Contract Code Differencing}
The analysis of upgrade changes involves systematically comparing the old and new versions of the code to identify changes. 
This process includes examining modifications, e.g., insertion and deletion, for smart contract code.
The well-known text differencing approach, adopted in \emph{git} systems, can only provide line-level file modification information. 
Although it is well-suited for code auditing task, it lacks the support for the analysis of fine-grained code modifications such as syntax-level changes for programs.
GumTree~\cite{falleri2014fine} is a popular syntax-level code differencing tool supporting  many programming languages including Java, Python, and Javascript.
GumTree computes four edit actions: insert, delete, update, and move, by comparing the abstract syntax tree (AST) of the new code with that of the old one.
Currently, GumTree has not supported the Solidity programming language.
To adapt GumTree for contract code differencing, 
we integrated a Solidity grammar parser\footnote{\url{https://github.com/JoranHonig/tree-sitter-solidity}} and developed the according tool called GumTree-Sol, in order to support our analysis of upgrade changes.
\paragraph{Results}
We ran GumTree-Sol on all the contract upgrades of 583 proxies and finally produced 2,598 comparison results where the comparisons are performed across code files of same names.
This is because the code repository of an implementation contract crawled from Etherscan could contain multiple files and mostly code files of the same names are comparable.

\Cref{tab:codechange} summarizes the code changes of these contract upgrades.
The first column shows the names of 16 syntax structures and the remaining columns show the count for the four edit actions, respectively.
In addition, the last row presents the overall numbers.
The syntax structures listed in~\cref{tab:codechange} reflect the various levels of code components of Solidity smart contracts~\cite{solidity}, from top-level \emph{contract, function} to low-level \emph{state variable, statement}.
We exclude code comment and variable declaration statement changes in~\cref{tab:codechange} as they usually have little effect of code semantics.
Note that function updates include only the insert, delete, and move of its parameters and modifier invocations, and the changes of function attributes, e.g., visibility, state mutability. 
The update of contract and modifier is also similar and we elide it to conserve space.


\Cref{tab:codechange} shows that overall contract upgrade changes normally meet that the insertion are created more while existing code structures are moved infrequently. 
The addition and removal of functions are the most frequent code changes.
Assembly statements are often be removed to reduce code complexity.
Conditional statements are deleted much more frequently than they are added, and loop statements are rarely introduced.
We believe such changes could optimize code performance, as both conditional checks and loop executions consume significant amounts of transaction gas\footnote{Gas is a unit of transaction fee on Ethereum.}.
Emit statements can create event logs to trace user actions. 
In~\cref{tab:codechange},
many emit statements are added probably for higher contract execution transparency.
There is a considerable number of changes on expression statements, which include a set of important features, e.g., the statements of \code{require}, \code{assert}, and internal or external function call.
Such changes could be used for bug fixes, functionality improvements, or functionality enhancements. 

\begin{tcolorbox}[size=title, opacityfill=0.1]
	\textbf{Answer to RQ1}: 
	There are a considerable number of code changes for contract upgrades.
	The addition and removal of functions are the two most common code changes.
\end{tcolorbox}

%% file: RQ2.tex
\begin{table}[t]\centering
	\caption{The intention taxonomy for contracts upgrading.}\label{tab:taxonomy-intention}
	\scriptsize
	\begin{tabular}{l|c|c}\toprule
		Goal &Category &Action \\\midrule

		\multirow{5}{*}{Preventive} &\multirow{2}{*}{Documentation Clarification (64\,\%)} &Comment Improvement (40\,\%)\\
		& &Comment Addition (48\,\%)\\
		\cmidrule{2-3}
		&\multirow{3}{*}{Code Optimization (50\,\%)} &Code Refactoring (17\,\%)\\
		& &Feature Removal (41\,\%)\\
		& &Redundancy Removal (30\,\%)\\
			\midrule
		\multirow{2}{*}{Adaptive} &Programming Language Change &Solidity Version Update (14\,\%)\\
			\cmidrule{2-3}
		&SDK Update &OpenZeppelin Libary Update (3\,\%)\\
			\midrule
		\multirow{9}{*}{Perfective} &\multirow{5}{*}{Usability Improvement (93\,\%)} &Functionality Addition (69\,\%)\\
		& &Functionality Update (78\,\%)\\
		& &Traceability Addition (27\,\%)\\
		& &Interoperability Addition (23\,\%)\\
		& &Exception Handling Enhancement (24\,\%)\\
		\cmidrule{2-3}
		&\multirow{4}{*}{Security Improvement (26\,\%)} &Access Control Mangement (16\,\%)\\
		& &Governance Update (23\,\%)\\
		& &Reentrancy Prevention Addition (15\,\%)\\
		& &Safe Operations Use (4\,\%)\\
		\midrule
		\multirow{3}{*}{Corrective} &\multirow{3}{*}{Bug Fix (25\,\%)} &Wrong Logic Correction (15\,\%)\\
		& &Code Typo Correction (11\,\%)\\
		& &Error Message Correction (9\,\%)\\
		\midrule
		\multirow{2}{*}{Others} &\multirow{2}{*}{Legal Aspects Change (12\,\%)} &License Update (4\,\%)\\
		& &Copyright Update (9\,\%)\\
		\bottomrule
	\end{tabular}
\end{table}


\subsection{RQ2: Upgrade Intention}
\label{sec:rq2}

\paragraph{Taxonomy}
%
To gain a comprehensive understanding of the intentions behind smart contract upgrades, we propose a
taxonomy in~\cref{tab:taxonomy-intention} that draws inspiration from traditional software change intention categorization while also
considering the unique characteristics of smart contracts. 
We adopt the well-accepted high-level change intentions firstly discussed
in~\cite{swanson1976dimensions} and further extended by~\cite{burd1999initial}, comprising an initial taxonomy
of four main change goals: 
\begin{itemize}[leftmargin=*]
	\item \emph{Preventive} includes changes aimed at improving code maintainability and clarity, such	as code optimization and comment improvements. 
	\item \emph{Adaptive} covers upgrades that adapt to changes in the programming language or external libraries.
	\item  \emph{Perfective} includes enhancements to features and security, such as adding new functionality, improving exception handling.
	\item  \emph{Corrective} covers bug fixes and error corrections.
\end{itemize}
Then we sampled a representative set consisting of 100 upgrade cases and manually inspected their code changes to
identify the underlying intentions. Based on this analysis on samples, we have further divided each main
category into subcategories to capture more granular intentions specific to smart contract upgrades.
At this time, we notice that some upgrades solely change license or copyright information. We add an
``Others'' type to include such changes besides the four aforementioned main categories. 
The final intention taxonomy concluded in~\cref{tab:taxonomy-intention} includes 21 specific upgrade actions  such as
updating the Solidity version, adding traceability, i.e., the addition of events. This
level of granularity allows for a more detailed analysis of the motivations behind smart contract
upgrades.
%
While the categorization may not be exhaustive due to the
ongoing contract evolution, our proposed taxonomy provides a basis for analyzing the currently available dataset. 
We hope this taxonomy can provide insights into the various intentions behind smart contract upgrades
and serve as a valuable tool for investigating the impact of upgrades on the overall functionality,
security, and maintainability of smart contracts.

\paragraph{Results}
To validate and refine our taxonomy based on the full dataset, we conducted a hybrid card sorting
process for all cases. Three authors independently reviewed the code changes and labeled all upgrade cases using the taxonomy, taking around 5 minutes per case, with a fourth author to break the tie in case of disagreements. Note that multiple labels can be assigned to a case. 
During this process, if needed, we proposed new actions as the refinement of the taxonomy.
Particularly, for wrong logic correction, we mainly examine if new security checks are added in upgrades. 

\begin{figure}[t]
	\centering
	\includegraphics[width=.55\columnwidth]{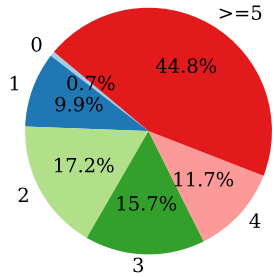}
	\caption{The distribution of upgrade changes with varied number of intentions.}
	\label{fig:upgradedistribution}
\end{figure}
Most smart contract upgrades are complicated.
\Cref{fig:upgradedistribution} shows the distribution of upgrade changes with varied number of intentions.
We observed a few upgrades (0.7\,\%) that have no actual changes according to our taxonomy.
These upgrades only did code formatting, identifier renaming, and expression rewriting.
Surprisingly,
44.8\,\% of the upgrade changes have at least five specific intentions, while only 9.9\,\% upgrades and 17.2 upgrades have one and two intentions, respectively.
As shown in \cref{tab:taxonomy-intention},
functionality update and addition are the two dominating upgrade intentions, accounting for 78\,\% and 69\,\% upgrades, respectively.
The least upgrade actions are OpenZeppelin library update and governance update only for 3\,\% upgrades.  
The standing intention category is usability improvement, followed by documentation clarification and code optimization.
In contrast, there are just a small proportion of upgrades for bug fixes and security improvement.
We list two upgrade cases for demonstration.
\Cref{fig:bugfixes} shows a bug fix example\footnote{\url{https://etherscan.io/address/0xee46ba5d4cc396d1ca4a7bf1a62f33b854e8831a\#code}} due to a simple programming mistake where "==" should be corrected as "=".
\Cref{fig:sou} shows a safe operation use example between the two implementations\footnote{\url{https://etherscan.io/address/0x653498a9e1c98eedaa9733d2d28617ff0e43dc18\#code}}\footnote{\url{https://etherscan.io/address/0xbcc10afafb91a736e47d435841b56eb9ffa5f402\#code }} that replaces \emph{transfer} with a safer version \emph{safeTransfer} that ensures the recipient can handle tokens, providing additional safety.

The uncovered complexity of contract upgrades makes it challenging to ensure the reliability and security of smart contracts. 
Removal of widely used features could cause incompatibility issues.
Unverified update of existing functionality may pose threats to smart contract correctness.
Adding functions into smart contracts seems straightforward to enrich its functionality, but chances are also high to introduce loopholes. 

%
%

\begin{figure}[t]
	\centering
	\small
	\begin{minted}[escapeinside=||,texcomments, linenos, breaklines,highlightlines={4}]{solidity}
   // and cursor.next is still in same bucket, move head to cursor.next
  if(infos[info.next].expiresAt.div(bucketStep) == bucket.div(bucketStep)) {
-        checkPoints[bucket].head == info.next;
+        checkPoints[bucket].head = info.next; // bugfix 
   } ...
	\end{minted}
	\caption{A bug fix example.}
	\label{fig:bugfixes}
\end{figure}

\begin{figure}[t]
	\centering
	\small
	\begin{minted}[escapeinside=||,texcomments, linenos, breaklines,breakanywhere,  highlightlines={1, 5}]{solidity}
+   using SafeERC20 for IERC20; 
   ...
    address addr = _state.collateralAssetList[i];
-   IERC20(addr).transfer(
+   IERC20(addr).safeTransfer(
	msg.sender,	
	actual.mul(IERC20(addr).balanceOf(address(this))).div(dollarTotalSupply)
  );
	\end{minted}
	\caption{A safe operation use example for token transfer.}
	\label{fig:sou}
\end{figure}

\begin{tcolorbox}[size=title, opacityfill=0.1]
	\textbf{Answer to RQ2}: 
	Smart contract upgrades are complicated.
	Functionality update and addition are the two dominating intentions. Most contract upgrades are for usability improvement, documentation clarification and code optimization.
\end{tcolorbox}


%% file: RQ3.tex
\begin{figure}[t]
	\centering
	\begin{subfigure}[b]{0.49\columnwidth}
		\small
		\begin{tabular}{ll}
			\toprule
			Change & \#Function \\
			\midrule
			\textcircled{1}-Removal & 3,614 \\
			\textcircled{2}-ParamerUpdate & 522 \\
			\textcircled{3}-ReturnChange & 198 \\
			\midrule 
			Overall & 4,334\\
			\bottomrule 
		\end{tabular}
		\caption{ABI breaking changes.}
		\label{fig:changefunction}
	\end{subfigure}
	\hfill
	\begin{subfigure}[b]{0.49\columnwidth}
		\includegraphics[width=\columnwidth]{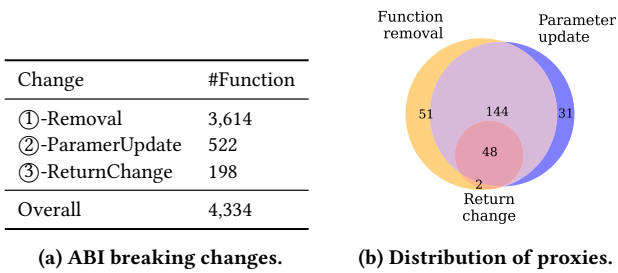}
		\caption{Distribution of proxies.}
		\label{fig:changeproxy}
	\end{subfigure}
	\caption{The statistics of ABI breaking changes.}
	\label{fig:breakingchanges}
\end{figure}
\begin{figure*}[t]
	\centering
	\includegraphics[width=.9\textwidth]{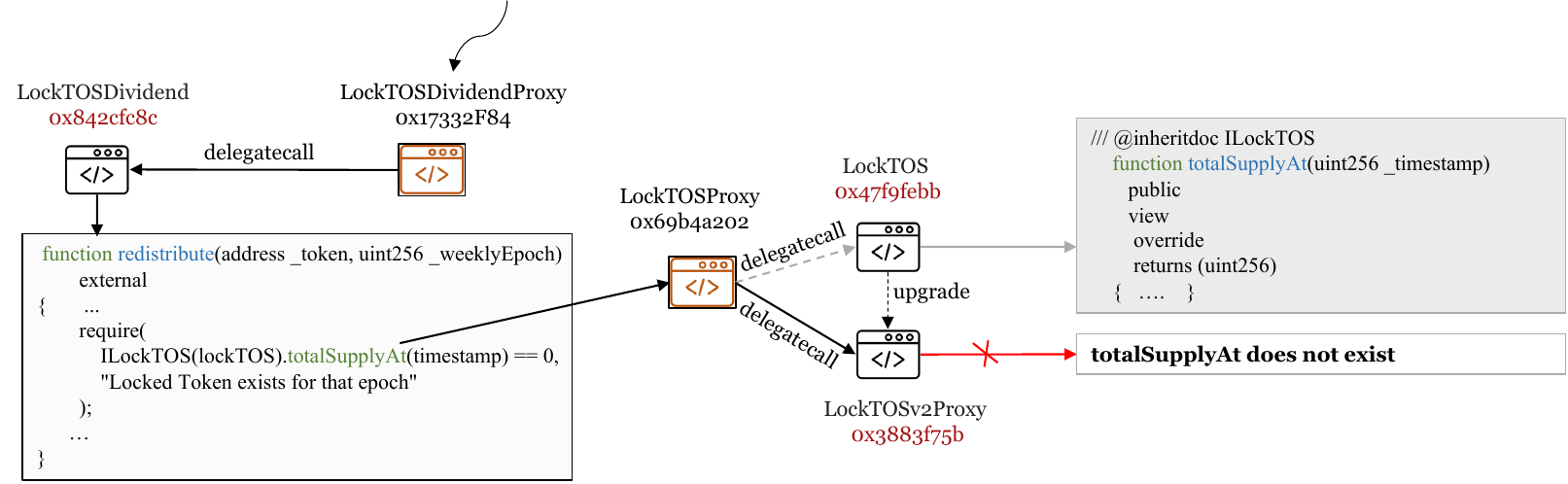}
	\caption{Calling to a deleted function ``totalSupplyAt'' after LockTOSProxy upgrading its implementations.}
	\label{fig:totalSupplyAt}
\end{figure*}
\begin{figure*}[t]
	\includegraphics[width=.9\textwidth]{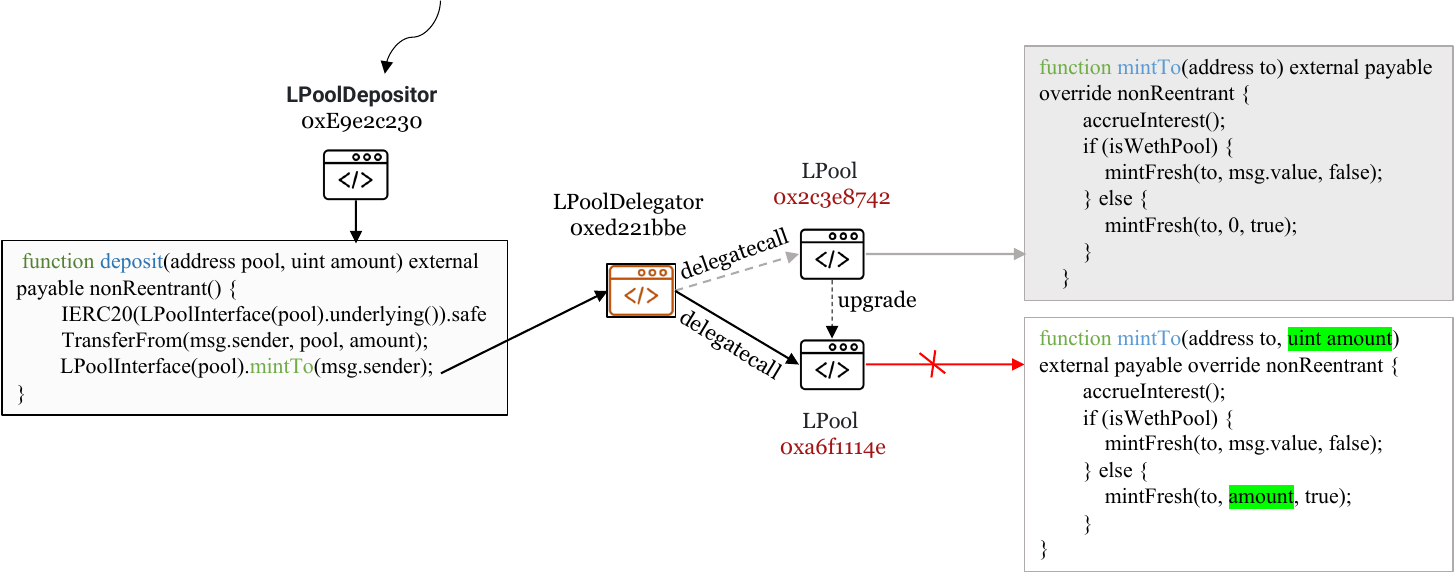}
	\caption{Incorrect parameters given to an updated function ``mintTo'' after LPoolDelegator upgrading its implementation.}
	\label{fig:mintTo}
\end{figure*}
\subsection{RQ3: Compatibility Issue}
\label{sec:rq3}
Upgrading smart contracts can introduce breaking changes to their Application Binary Interfaces (ABIs), leading to compatibility issues for other contracts/DApps built upon them, as well as for regular users unaware of these ABI upgrades. 
Generally, a breaking change refers to the code change that dissatisfies the full backward compatibility between the current and previous version of programs, causing functionality inequivalence~\cite{raemaekers2017semantic}.
However, semantic equivalence check between programs is in general undecidable, especially for smart contracts that have many unspecified behaviors, making it challenging to have universal equivalence criteria.
For instance, developers often mix the use of \code{require} and \code{assert} statements in smart contracts to perform security checks, where their difference is very subtle, i.e., varied transaction cost when the security check fails.
In this work, we focus on the client-specific compatibility related to ABI changes, which is often called \emph{binary compatibility}~\cite{KindsofCompatibility}, and leave the semantic equivalence check for future work.
Smart contracts have to be compiled from source code into binary bytecode runnable on the blockchain.  
ABI changes are \emph{binary compatible} with preexisting usages if the usages that previously linked without error will continue to link without error.

\paragraph{Compatibility Checking}
To perform such compatibility checking, 
we studied all the pairs of the current and previous version of contract ABIs and explored the real-world broken usages caused by the ABI breaking changes.
We crawled the ABI specifications from Etherscan and eventually succeeded in obtaining ABI files for 834 implementation contracts.
We categorize ABI breaking changes into the following three main classes.
\begin{itemize}[leftmargin=*]
	\item \textcircled{1}-Removal. Functions in the previous contract ABI are deleted in the current version of ABI.
	\item \textcircled{2}-ParameterUpdate. The number, order and types of function input parameters in the previous contract ABI are modified in the current version of ABI. 
	\item \textcircled{3}-ReturnChange. The types of function return data in the previous contract ABI  are changed in the current version of ABI. 
\end{itemize}

\paragraph{Results}
\Cref{fig:breakingchanges} presents the number of functions having breaking changes and the distribution of the corresponding proxy contracts.
\Cref{fig:changefunction} shows that there are 4,334 ABI breaking changes for contract upgrades.
\textcircled{1}-Removal accounts for 83\,\% ABI breaking changes.
The least breaking changes are about \textcircled{3}-ReturnChange, occupying only 4.6\,\%.
 \Cref{fig:changeproxy} shows that there are 276 proxies having ABI breaking changes in its upgrades, where 245 proxies contain \textcircled{1}-Removal, followed by 223 proxies for \textcircled{2}-ParameterUpdate, and 50 proxies for \textcircled{3}-ReturnChange.
 In terms of their overlapping, 48 proxies have all the three kinds of ABI breaking changes, while the majority of proxies (144) have both \textcircled{1}-Removal and \textcircled{2}-ParameterUpdate.
 
 We conducted a thorough scan on all the historic transactions to identify broken usages due to these ABI breaking changes.
 Specifically, we collected the transactions between Janurary 1, 2018 and Janurary 1, 2024 that are related to the proxy contracts from the Ethereum dataset public in BigQuery\footnote{https://cloud.google.com/blog/products/data-analytics/ethereum-bigquery-public-dataset-smart-contract-analytics}. 
 The results show that there are in total 12,941,660 transactions involving 32,314,209 usage calls to the studied 583 proxy contracts.
To this end, we found 498 transactions containing broken usage calls to eight proxies for \textcircled{1}-Removal and another 86 transactions containing 
 broken usage calls  to nine proxies for \textcircled{2}-ParameterUpdate, but zero broken usage calls detected for \textcircled{3}-ReturnChange.
 We highlight the impact of ABI breaking changes with the following two broken usage cases.

\paragraph{Call to a deleted function}
\Cref{fig:totalSupplyAt} shows a broken usage caused by~\textcircled{1}-Removal, i.e., calling to a deleted function \code{totalSupplyAt} after contract upgrade.
LockTOS is a time lock smart contract application that allows users to deposit a certain amount of tokens where users can only withdraw his money after the previously configured time lock expires.
Users receive dividends based on the staked period of their money through the function \code{redistribute} of contract \code{LockTOSDividend}.
As shown in~\cref{fig:totalSupplyAt},
LockTOSProxy is a proxy contract that have upgraded its implementation from the original LockTOS to LockTOSv2Proxy, a diamond proxy contract that could have multiple implementation contracts for different functionality.
However, \code{totalSupplyAt} is absent in the contract ABI of LockTOSv2Proxy and its corresponding implementations, making \code{redistribute} fail consistently due to the external call to the non-existent \code{totalSupplyAt}.  
	There are two practical fixes:
	(1) LockTOSDividend can be upgraded by its proxy contract LockTOSDividendProxy to discard functions related to the external call of totalSupplyAt;
	(2) LockTOSProxy can re-upgrade its implementations to restore totalSupplyAt function
%

\paragraph{Incorrect parameters provided}
\Cref{fig:mintTo} shows a broken usage caused by \textcircled{2}-ParameterUpdate, i.e., incorrect parameters given to an updated function \code{mintTo} after contract upgrade.
LPool is a liquidity pool application in DeFi area.
When users deposit a certain amount of tokens in this pool, they will accordingly receive liquidity token as the so-called stakes.
The contract \code{LPoolDepositor} is responsible to implement the aforementioned business in a function \code{deposit}, which first transfers tokens from users to the pool and then rewards users with liquidity tokens minted by an external call to the \code{mintTo} function of proxy contract \code{LPoolDelegator}.
However, LPoolDelegator has upgraded its implementation and the current version (0xa6f1114e) introduced an additional parameter ``amount'' (colored in green) in \code{mintTo}.
Therefore, \code{deposit} always fails to execute successfully due to the unreachable call to the modified mintTo after this upgrade. 
As LPoolDepositor is not an upgradable smart contract,
to fix this incompatibility issue, LPoolDelegator should upgrade it implementation contract to overload mintTo function to achieve backward compatibility.


To avoid compatibility issues during contract upgrades, developers should minimize unnecessary modifications to existing interface functions.
Furthermore, client users of upgradable contracts must monitor implementation changes to swiftly adapt to any arising compatibility issues. If another client contract uses the outdated ABI of an upgradable contract, its developers must address this by redeploying or upgrading the contract. Offchain users or DApps must also vigilantly track ABI-breaking changes and adjust their usage practices accordingly.

\begin{tcolorbox}[size=title, opacityfill=0.1]
	\textbf{Answer to RQ3}: 
	Compatibility issues exist in smart contract upgrades. There are 4,334 ABI breaking changes involving more than 276 proxies' upgrades, which has a significant impact on real-world contract usages.
\end{tcolorbox}

%% file: RQ4.tex
\subsection{RQ4. Security Issue}\label{sec:rq4}

There lacks a universal standard for correctly implementing upgradable smart contracts, leaving
rooms for many critical security vulnerabilities.
\begin{figure}
	\centering
	\includegraphics[width=.9\columnwidth]{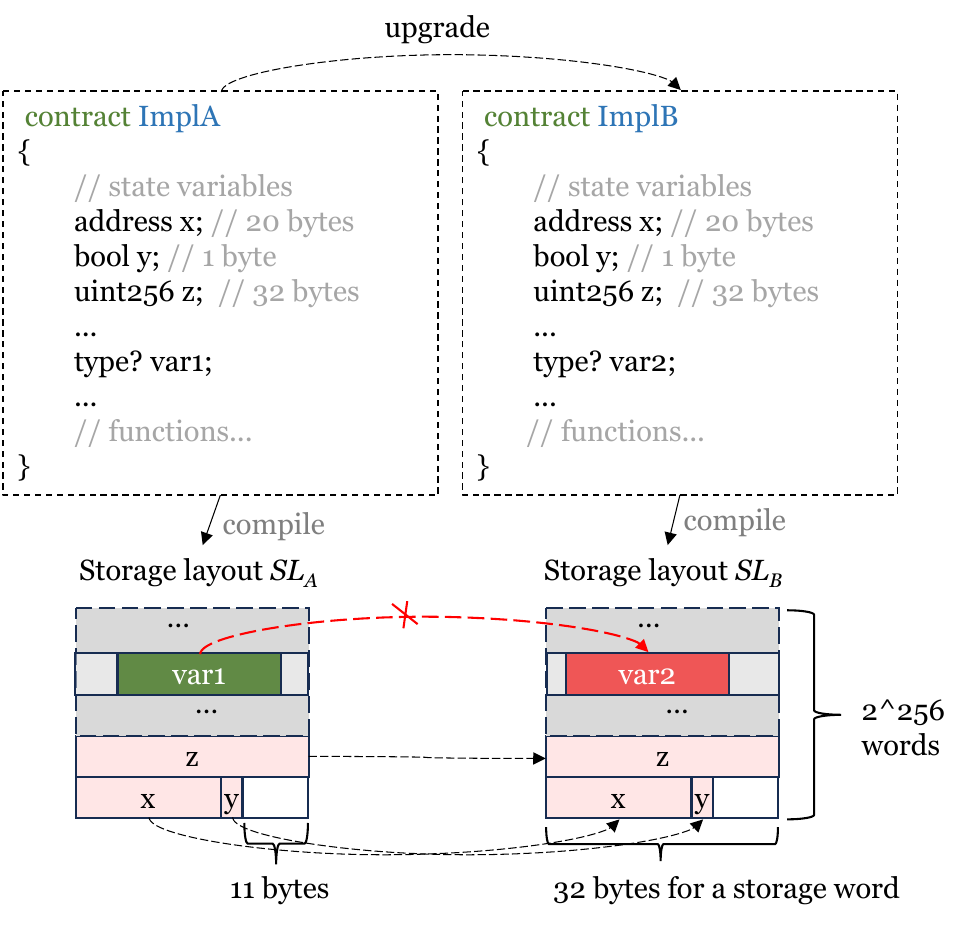}
	\caption{Storage collision problem.}
	\label{fig:collision}
\end{figure}
\input{./figure/storageCollision}

\paragraph{Storage Collisions}
Storage collisions are the most common security issues for upgradable smart contracts and have been
previously analyzed~\cite{OpenZeppelin,bodell2023proxy}.
\Cref{fig:collision} illustrates the storage collision problem for a contract upgrade from an
implementation \code{ImplA} to another \code{ImplB}.
The state variables of a smart contract are stored at different slots in blockchain storage
according to the compilation rules~\cite{solidity}.
A word on Ethereum occupies 32 bytes and the storage space cannot exceed $2^{256}$ words.
In~\cref{fig:collision}, the variables \code{x} and \code{y} are packed into the first storage word
because their total bytes occupied are less than 32 bytes, but leaving 11 bytes untouched, because
the variable \code{z} occupies 32 bytes and is aligned to a fresh storage word.
A storage collision occurs when two different variables, e.g., \code{var1} and \code{var2},
reference to the same storage slot, resulting in unintended behaviors.

We used Slither~\cite{feist2019slither} to compile the implementation contracts and finally obtained storage layout specifications for 498 contracts.
For each upgrade, we compared the storage layout of the previous implementation to that of the
current implementation if applicable.
Specifically, we checked whether all state variables of the previous contracts exist at the same storage location in the current ones.
To achieve fuzzy match between state variables, we allow small variations on the variable names in
case name changes are introduced during the upgrades.
Eventually, we detected 41 storage collision cases.
After manual confirmation, 5 cases are false positives and the remaining 36 are true positives,
achieving a precision rate of 87\,\%.

\Cref{fig:storageCollision} and \cref{fig:storagepractice} show a true positive and false positive,
respectively.
\Cref{fig:storageCollision} shows storage collision caused by the addition of the enum variable \code{depositMode}.
A storage word on Ethereum is 32 bytes.
The size of \code{address} is 20 bytes while \code{enum} and \code{bool} occupy only one byte.
As said in the code comment (Line~\ref{depositMode:comment}),
\code{vaultProxy}, \code{depositMode} and three following variables
(Lines~\ref{depositMode:start}-\ref{depositMode:end}) are packed into the same storage word, which
will not tamper the existing value of the variable \code{redemptionAsset}
(Line~\ref{depositMode:redemptionAsset}).
However, \code{depositMode} collides with the previously existing value of
\code{useDepositApprovals}, breaking the integrity of the contract state.

The false positive shown in \cref{fig:storagepractice} is indeed a good practice for state variable
abolishment and addition.
The key is to ensure the continuous storage space of smart contracts always remains constant during upgrading.
To abolish state variables that are no longer in use, it is safe to deprecate it with special tags,
e.g., prefix \code{\_deprecated\_}
(Lines~\ref{storagepractice:deprecate:start}-\ref{storagepractice:deprecate:end}) rather than
a simple deletion.
To introduce new state variables, it is recommended to reserve a large continuous storage space,
e.g., a dummy integer array \code{\_reserved} of 100 storage words
(Line~\ref{storagepractice:reserve}), in the early contract design.
Therefore, new state variables, e.g., \code{harversterAddress} and \code{rewardTokenAddresses}
(Lines~\ref{storagepractice:assign:start}-\ref{storagepractice:assign:end}), can be safely enqueued
into the previously reserved space.
\paragraph{Initialization Risk}
Developers may forget to initialize a contract, leading to serious consequences, as in the case of Wormhole's uninitialized proxy~\cite{wormhole}.
Initialization risks for upgradable smart contracts include unprotected initialization, where lack of protection allows multiple calls to initialization functions, enabling attackers to reset the contract state.
In our dataset, we found that 75\,\% (732) implementation contracts have initialization functions
with names to be either \code{initialize} or \code{init}.
We tested these contracts and their related proxy contracts on a forked blockchain via Infura API
to trigger their initialization functions.
Our detection results show that 5 proxies and 59 implementation contracts can be successfully initialized by us, with no false positives.
To avoid ethical implications, we took an implementation contract
\code{Deployer1}\footnote{\url{https://etherscan.io/address/0xa134da59c6e75566e274937d09fe4f745731c030\#code}}
 abandoned three years ago as a case study.
This contract has already been uninitialized and its \code{initialize} function can still be
arbitrarily called.


Other vulnerabilities in upgradable contracts occur when bugs are introduced in the contract's
logic during upgrades~\cite{euler, Pikehack}, potentially leading to unintended behaviors or
exploitations.
There are no established methods to detect logic vulnerabilities for contract upgrades, hence
rigorous testing and audits are crucial for developers to mitigate the risk of such vulnerabilities.

\begin{tcolorbox}[size=title, opacityfill=0.1]
	\textbf{Answer to RQ4}: There are many storage collisions during contract upgrades and developers
	should follow best practices. The proper initialization of upgradable smart contracts is also
	important. Developers shall check the initialization status for both proxy and implementation
	contracts during upgrades.

\end{tcolorbox}

%% file: figure/storageCollision.tex
\begin{figure}[t]
	\centering
	\small
	\begin{minted}[escapeinside=||,texcomments, linenos, breaklines,highlightlines={9}]{solidity}
abstract contract GatedRedemptionQueueSharesWrapperLibBase1{
...
+  enum DepositMode {
+	Direct,
+	Request
+  }
+   // Packing vaultProxy with depositMode and useDepositApprovals makes deposits slightly cheaper  \label{depositMode:comment} 
   address internal vaultProxy; |\label{depositMode:start}|
+  DepositMode internal depositMode; 
   bool internal useDepositApprovals;
   bool internal useRedemptionApprovals;
   bool internal useTransferApprovals; |\label{depositMode:end}|
   address internal redemptionAsset; |\label{depositMode:redemptionAsset}|
}
	\end{minted}
	\caption{A storage collision case for adding  ``depositMode''.}
	\label{fig:storageCollision}
\end{figure}

\begin{figure}[t]
	\centering
	\small
	\begin{minted}[escapeinside =||,texcomments, linenos, breaklines,highlightlines={5,6,11-16}]{solidity}
pragma solidity ^0.8.0;
abstract contract InitializableAbstractStrategy {
- // Reward token address
- address public rewardTokenAddress;} 
+ // Deprecated: Reward token address |\label{storagepractice:deprecate:start}|
+ address public _deprecated_rewardTokenAddress;  |\label{storagepractice:deprecate:end}|

-  // Reserved for future expansion
- int256[100] private _reserved; |\label{storagepractice:reserve}|

+ // Address of the one address allowed to collect reward tokens |\label{storagepractice:assign:start}|
+ address public harvesterAddress;
+ // Reward token addresses
+ address[] public rewardTokenAddresses;
+  // Reserved for future expansion  \label{storagepractice:assign:end}
+ int256[98] private _reserved;
}
	\end{minted}
	\caption{A false positive but with good state variable abandon and extension practice.}
	\label{fig:storagepractice}
\end{figure}

%% file: discuss.tex
\section{Discussion}
\label{sec:discussion}

\subsection{Implications}

\noindent\textbf{Validating contract upgrades.}
Upgradable smart contracts are valuable to fix  bugs and evolve contract functionality transparently.
However, upgrades of smart contracts can introduce breaking changes, affecting normal user
experiences.
Additionally, security vulnerabilities could arise in new contract versions.
To validate contract upgrades, we anticipate the following efforts from users, developers, and academia.
\begin{itemize}[leftmargin=*]
	\item For users, they should be aware of smart contract upgrades and the potential compatibility
	issues.
	Specifically, users could subscribe to upgrade activities, e.g., periodically checking the change of the
	logic implementation by calling the ``implementation()'' function.
	They could also employ \tool to automatically identify ABI breaking changes to help inspect
	compatibility issues.
	\item {For developers, they should upgrade smart contracts with full backward compatibility by function overloading. Developers  also need to check storage layout change and initialization status of either proxy and implementation contracts. It is recommended to deprecate state variables without deletion and to reserve a fixed-size storage for state variables expansion.}
	\item {For academia, our empirical study on smart contract upgrades opens a new area of research in automated software engineering. Developing a fully automated and comprehensive approach for smart contract upgrades remains challenging due to the complexity of the upgrade changes involved.
	Additionally, contract design requirements and runtime environments may evolve over time, which make it difficult to derive reliable security models.
}
\end{itemize}

\noindent\textbf{Managing contract versions.}
Explicitly indicating change intentions of smart contract upgrades is useful for effectively
managing a potentially large number of implementation contract versions.
The comprehensive upgrade intention taxonomy proposed in our work is based on a systematic study
of the intentions behind real-world smart contract upgrades, which could serve as a
reference to standardize intention specifications.
On the other hand, the existing software versioning mechanisms, e.g., Git and SVN, are no longer
effective for smart contract upgrades that are initiated on blockchains.
New versioning systems, which not only track code differences but also record up-to-date meta
information pertinent to the deployment environments, should be developed to support better
transparency and traceability.

%

\subsection{Threats to Validity}
\label{sec:threats}
\paragraph{Internal Validity}
There is no ground truth about the intentions for contract upgrades.
To mitigate this problem, we first design a comprehensive intention taxonomy derived from established works and also consider the unique characteristics of smart contracts.
To avoid the bias of manual intention analysis, three authors independently labeled these upgrades and a fourth author resolved the disagreements.

\paragraph{External Validity}
The upgradable smart contracts studied could be limited, and our findings may not generalize to other upgradable smart contracts.
Proxies are the most common form of upgradable smart contracts.
We systematically collected a large-scale open source proxy contracts from different blockchains and found 72.7\% proxy contracts include the required ``implementation()'' interface function and they have long-standing usages.
Therefore, we believe these proxy contracts are representatives and our research findings could be applicable to other kinds of contract upgrades.

%% file: related.tex
\section{Related Work}
\label{sec:related}
\subsection{Smart Contract Upgrade} 
Since smart contracts are immutable once deployed,
upgrades provide the only way to perform post-deployment smart contract maintenance~\cite{metcalfe2020ethereum}, either to patch bugs~\cite{rodler2021evmpatch}, to add functionality, or to improve contract features.
There are many upgradability patterns.
\citeauthor{bodell2023proxy}~\cite{bodell2023proxy} summarized eleven unique proxy-based upgradability patterns, e.g., inherited storage,  external storage, and singleton etc.
Apart from proxy patterns,  \citeauthor{xiaofan2024}~\cite{xiaofan2024} uncovered more upgradability patterns including data separation, strategy pattern, and metamorphic contracts using new opcode \code{CREATE2} for upgrading.
Instead of upgradability analysis, our work delves into characterizing real-world upgrades to understand how developers exploit upgradability.

The security and reliability of upgradable smart contracts have garnered considerable attention. 
The issues on insufficient compatibiltiy checks of upgradability-related functions, e.g., \code{setter} for implementation contracts, has been studied and investigated in~\cite{bodell2023proxy,xiaofan2024}.
In contrast, the compatibility issues studied in our works prioritize attentions to the benign usage of upgradable smart contracts.
\citeauthor{chen2020finding}~\cite{chen2020finding} proposed to find four security issues, e.g., unmatched ERC20 contract, by comparing historic contract versions used.
\citeauthor{galimullin2022coalition}~\cite{galimullin2022coalition} applied a temporal logic~\cite{galimullin2021dynamic} for specification and verification of smart contract upgrades and employed model checking techniques to discover upgrade-related issues.
Similarly, \citeauthor{sorensen2024towards}~\cite{sorensen2024towards} introduced a formal framework, ConCert, which is based on Coq~\cite{huet1997coq} to specify and verify smart contract upgrades to avoid faulty updates.
Furthermore, to prevent upgrade issues, \citeauthor{antonino2022specification}~\cite{antonino2022specification} raised a new paradigm of ``specification is law'' for contract upgrades and proposed a novel upgrading framework centered around an offchain trust deployer to enforce the conformance of given formal specifications prior to contract upgrades. 
\tool aligns with the aforementioned upgrade analyses, but focuses on the detection of two common vulnerabilities, \emph{storage collisions}, and \emph{initialization risks}, which are critical to the data integrity of smart contracts.

\subsection{Traditional Software Evolution}
Our research on contract evolution is also closely connected to the wider domain of software evolution,
especially for API libraries in open source area.
 \citeauthor{li26}~\cite{li26} summarized 16 fine-grained web service API change patterns such as \emph{decrease or increase number of parameters}, \emph{change type of parameters}, and \emph{change format of parameters}, which are similar to the changes included as \emph{ParameterUpdate} in our work. 
\citeauthor{zhang10}~\cite{zhang10} included removal and addition of class field as two API evolution patterns for Python frameworks but considered only field removal as one of the breaking changes.
In contrast, our studied changes, e.g., addition and reordering,  to the state variables of smart contracts are probably the severe breaking changes, leading to storage collision problems.

API evolution also causes issues.
\citeauthor{zhang23}~\cite{zhang23} studied TensorFlow applications and found the API changes account for 40.9\,\% issues.  	
To detect compatibility issues.
\citeauthor{zhang10}~\cite{zhang10} proposed a tool called PYCOMPAT by searching usages of evolved APIs for the detection of API renaming/relocating and parameter renaming issues.
Moreover, to mitigate the side effect of API evolution,
\citeauthor{wang16}~\cite{wang16} proposed a declarative language to facilitate the migration of Java programs between different API versions.
In this work,
we make the initial step to uncover upgrade issues in real-world smart contracts which we think could shed light on upgrade issue analysis and migration of smart contracts.
%

%% file: conclusion.tex
\section{Conclusion and Future Work}
\label{sec:conclusion}
In this paper, we propose a comprehensive analysis framework called \tool for smart contract upgrades.
We investigate 57,118 proxy contracts from different blockchains.
Our empirical study uncovers the characteristics of upgrade changes and intentions for upgradable smart contracts.
In addition, we find a considerable number of client-specific compatibility issues and confirm the substantial existence of storage collisions and initialization risks.
For future work, we plan to support semantic equivalence check between contracts in \tool to identify logic bugs incurred by contract upgrades.